\documentclass[11pt,conference]{IEEEtran}
\usepackage{blindtext, graphicx}

\usepackage{balance}  
\usepackage{fixltx2e}
\usepackage{url}
\let\labelindent\relax
\usepackage{enumitem}
\usepackage{comment} 
\usepackage{graphicx}
\usepackage{balance}  
\usepackage{url}
\usepackage{listings}
\usepackage[acronym,toc,shortcuts]{glossaries}
\usepackage{epsfig}
\usepackage{amssymb}
\usepackage{amsthm}
\usepackage{caption}
\usepackage{subcaption}
\usepackage{siunitx}
\usepackage{soul}
\usepackage{csquotes}
\usepackage{bbm}
\usepackage{dsfont}
\usepackage{xcolor}
\usepackage{enumitem}
\usepackage{adjustbox}
\usepackage{array}
\usepackage{multirow}
\usepackage{fancyhdr}
\usepackage[T1]{fontenc}
\usepackage{listings}
\usepackage{tikz}
\usepackage[official]{eurosym}
\usetikzlibrary{positioning}

\definecolor{delim}{RGB}{20,105,176}

\lstdefinelanguage{json}{
    basicstyle=\normalfont\ttfamily,
    numberstyle=\scriptsize,
    stepnumber=1,
    numbersep=8pt,
    showstringspaces=false,
    breaklines=true,
    frame=lines,
}

\usepackage{colortbl} 

\graphicspath{{figures/}}

\makeglossaries
\newacronym{3GPP}{3GPP}{The 3rd Generation Partnership Project }
\newacronym{5G}{5G}{Fifth Generation}
\newacronym{AAA}{AAA}{Authentication, Authorization and Accounting}
\newacronym{AI}{AI}{Artificial Intelligence}
\newacronym{AES}{AES}{Advanced Encryption Cluster}
\newacronym{AoI}{AoI}{Age of information}
\newacronym{AP}{AP}{Access Point}
\newacronym{API}{API}{Application Programming Interface}
\newacronym{APN}{APN}{Access Point Name}
\newacronym{BS}{BS}{Base Station}
\newacronym{BER}{BER}{Bit Error Rate}
\newacronym{BSSID}{BSSID}{Basic Service Set Identification}
\newacronym{CAT}{CAT}{Capacity-Aware TOPSIS}
\newacronym{CEP}{CEP}{Complex Event Processing}
\newacronym{CELL-ID}{CELL-ID}{cell identification ID}
\newacronym{CGI}{CGI}{Cell Global Identification}
\newacronym{C2}{C2}{Command and Control}
\newacronym{CI}{CI}{Confidence Interval}
\newacronym{CLSM}{CLSM}{Closed loop spatial multiplexing}
\newacronym{CQI}{CQI}{Channel Quality Indicator}
\newacronym{CN}{CN}{core network}
\newacronym{CNN}{CNN}{Convolutional Neural Networks}
\newacronym{CL}{CL}{Closed-Loop}
\newacronym{CoMP}{CoMP}{coordinated multi-point}
\newacronym{CPU}{CPU}{Central Processing Unit}
\newacronym{CS}{CS}{central scheduler}
\newacronym{CSI}{CSI}{Channel Status Information}
\newacronym{eNB}{eNB}{evolved Node-B}
\newacronym{DaaS}{DaaS}{Data as a Service}
\newacronym{DL}{DL}{Downlink}
\newacronym{DLT}{DLT}{Distributed Ledger Technology}
\newacronym{DMM}{DMM}{Distributed Mobility Management}
\newacronym{DST}{DST}{Dempster-Shafer Theory}
\newacronym{ECA}{ECA}{Event-Condition-Action}
\newacronym{ECC}{ECC}{Elliptic-curve cryptography}
\newacronym{eNodeB}{eNodeB}{evolved Node-B}
\newacronym{E-RAB}{E-RAB}{E-UTRAN Radio Access Bearer}
\usepackage{algorithm}
\usepackage{algorithmic}
\newacronym{ETSI}{ETSI}{European Telecommunications Standards Institute}
\newacronym{FDD}{FDD}{Frequency Division Duplexing }
\newacronym{FEM}{FEM}{Flow Extraction Manager}
\newacronym{GDPR}{GDPR}{General Data Protection Regulation}
\newacronym{GGSN}{GGSN}{Gateway GPRS Support Node}
\newacronym{GPRS}{GPRS}{General packet radio service}
\newacronym{GTP}{GTP}{GPRS Tunneling Protocol}
\newacronym{HAPS}{HAPS}{High-Altitude Platform Station}
\newacronym{HetNet}{HetNet}{heterogeneous network}
\newacronym{HSS}{HSS}{Home Subscriber Station}
\newacronym{HTTP}{HTTP}{Hypertext Transfer Protocol}
\newacronym{HDFS}{HDFS}{Hadoop Distributed File Cluster}
\newacronym{HiveQL}{HiveQL}{Hive Query language}
\newacronym{HSPA}{HSPA}{High Speed Packet Access}
\newacronym{HUMINT}{HUMINT}{Human Intelligence}
\newacronym{IBLER}{IBLER}{Initial Block Error Rate}
\newacronym{ICIC}{ICIC}{inter-cell interference coordination}
\newacronym{ICN}{ICN}{information-centric network}
\newacronym{IEEE}{IEEE}{Institute of Electrical and Electronics Engineers}
\newacronym{IETF}{IETF}{Internet Engineering Task Force}
\newacronym{IMINT}{IMINT}{Imagery Intelligence}
\newacronym{IMSI}{IMSI}{International Mobile Subscriber Identity}
\newacronym{IMEI}{IMEI}{International Mobile Station Equipment Identity}
\newacronym{IMS}{IMS}{IP Multimedia SubCluster}
\newacronym{ICMP}{ICMP}{Internet Control Message Protocol}
\newacronym{IoT}{IoT}{Internet of Things}
\newacronym{InP}{InP}{Infrastructure Provider}
\newacronym{Autonomous Driving}{Autonomous Driving}{Intelligence, Surveillance and Reconnaissance}
\newacronym{ITU}{ITU}{International Telecommunication Union}
\newacronym{IT}{IT}{Information Technology}
\newacronym{GBR}{GBR}{Guaranteed Bit Rate}
\newacronym{GPS}{GPS}{Global Positioning Cluster}
\newacronym{GLUE}{GLUE}{General Language Understanding Evaluation}
\newacronym{JSON}{JSON}{JavaScript Object Notation}
\newacronym{KPI}{KPI}{Key Performance Indicator}
\newacronym{LA}{LA}{Location Area}
\newacronym{LAC}{LAC}{location area code}
\newacronym{LMA}{LMA}{Local Mobility Anchor}
\newacronym{LTE}{LTE}{long term evolution}
\newacronym{MADM}{MADM}{Multiple Attribute Decision Making}
\newacronym{MCC}{MCC}{Mobile Country Code}
\newacronym{MCS}{MCS}{Modulation Coding Scheme}
\newacronym{MNC}{MNC}{Mobile Network Code}
\newacronym{MIMO}{MIMO}{multiple-input multiple-output}
\newacronym{MAG}{MAG}{Mobile Access Gateway}
\newacronym{MAAR}{MAAR}{Mobility Anchor and Access Router}
\newacronym{ML}{ML}{Machine Learning}
\newacronym{MME}{MME}{Mobility Management Entity}
\newacronym{MN}{MN}{Mobile Node}
\newacronym{MNO}{MNO}{Mobile Network Operator}
\newacronym{MSISDN}{MSISDN}{Mobile Station International Subscriber Directory Number}
\newacronym{NBI}{NBI}{NorthBound Interface}
\newacronym{NIST}{NIST}{National  Institute  of  Standards  and Technology}
\newacronym{NLP}{NLP}{Natural Language Processing}
\newacronym{NLU}{NLU}{Natural Language Understanding}
\newacronym{NOMA}{NOMA}{Non-Orthogonal Multiple Access}
\newacronym{NoSQL}{NoSQL}{Not Only SQL}
\newacronym{NR}{NR}{New Radio}
\newacronym{NTN}{NTN}{Non-Terrestrial Networks}
\newacronym{QoS}{QoS}{quality-of-service}
\newacronym{QoE}{QoE}{quality-of-experience}
\newacronym{OAM}{OAM}{Operation, Administration and Management}
\newacronym{ONF}{ONF}{Open Networking Foundation}
\newacronym{ONOS}{ONOS}{Open Network Operating Cluster}
\newacronym{OS}{OS}{operating Cluster}
\newacronym{OL}{OL}{Open-Loop}
\newacronym{PDN}{PDN}{packet data network}
\newacronym{PF}{PF}{Proportional Fair}
\newacronym{P-GW}{P-GW}{packet gateway}
\newacronym{PDP}{PDP}{Packet Data Protocol}
\newacronym{PHY}{PHY}{physical layer}
\newacronym{PMIPv6}{PMIPv6}{Proxy Mobile IPv6}
\newacronym{PMI}{PMI}{Precoding Matrix Index}
\newacronym{PoW}{PoW}{Proof-of-Work}
\newacronym{PRB}{PRB}{Physical Resource Block}
\newacronym{PUSCH}{PUSCH}{Physical Uplink Shared Channel}
\newacronym{QAM}{QAM}{Quadrature amplitude modulation}
\newacronym{QRSA}{QRSA}{Quantum Resistant Security Algorithm}
\newacronym{QCI}{QCI}{QoS Class Identifier}
\newacronym{RA}{RA}{Routing Area}
\newacronym{RB}{RB}{Resource Block}
\newacronym{RI}{RI}{Rank Indicator}
\newacronym{RAN}{RAN}{radio access network}
\newacronym{RFC}{RFC}{Request for Comment}
\newacronym{RIS}{RIS}{Reconfigurable Intelligent Surfaces}
\newacronym{RRC}{RRC}{Radio Resource Control}
\newacronym{RNC}{RNC}{radio network controller}
\newacronym{RNN}{RNN}{Recurrent Neural Networks}
\newacronym{RSA}{RSA}{Rivest-Shamir-Adleman}
\newacronym{RSSI}{RSSI}{Received Signal Strength Indicator}
\newacronym{RSRP}{RSRP}{Reference Signal Received Power}
\newacronym{OTT}{OTT}{over-the-top}
\newacronym{SA}{SA}{Stand Alone}
\newacronym{SAC}{SAC}{service area code}
\newacronym{SCMA}{SCMA}{Sparse Code Multiple Access}
\newacronym{SLA}{SLA}{Service Level Agreement }
\newacronym{SDN}{SDN}{Software Defined Networking}
\newacronym{SDO}{SDO}{Standards Developing Organization}
\newacronym{SFN}{SFN}{Single Frequency Network}
\newacronym{S-GW}{S-GW}{serving gateway}
\newacronym{SINR}{SINR}{signal-to-interference-plus-noise ratio}
\newacronym{SIGINT}{SIGINT}{Signals Intelligence}
\newacronym{SGSN}{SGSN}{Serving GPRS Support Node}
\newacronym{SP}{SP}{Service Provider}
\newacronym{SSID}{SSID}{Service Set Identification}
\newacronym{SVD}{SVD}{singular value decomposition}
\newacronym{TCP}{TCP}{transport control protocol}
\newacronym{TDD}{TDD}{Time Division Duplexing}
\newacronym{TM}{TM}{transmission mode}
\newacronym{TEID}{TEID}{tunnel endpoint identifier}
\newacronym{UAV}{UAV}{Unmanned Aerial Vehicle}
\newacronym{UDN}{UDN}{Ultra Dense Network}
\newacronym{UMTS}{UMTS}{Universal Mobile Telecommunications Service} 
\newacronym{UE}{UE}{user equipment}
\newacronym{UL}{UL}{Uplink}
\newacronym{UDP}{UDP}{User Datagram Protocol}
\newacronym{V2V}{V2V}{vehicle-to-vehicle}
\newacronym{V2X}{V2X}{Vehicle-to-Everything}
\newacronym{V2I}{V2I}{Vehicle-to-Infrastructure}
\newacronym{VM}{VM}{Virtual Machine}
\newacronym{VNF}{VNF}{Virtual Network Function}
\newacronym{WiFi}{WiFi}{Wireless Fidelity}
\newacronym{XAI}{XAI}{Explainable AI}
\newacronym{WLAN}{WLAN}{Wireless Local Area Network}

\IEEEoverridecommandlockouts

\begin{document}
%

\title{HAPS-RIS-assisted IoT Networks for Disaster Recovery and Emergency Response: Architecture, Application Scenarios, and Open Challenges\vspace{-1cm}}


\author{ Bilal Karaman,  Ilhan Basturk, \IEEEmembership{Senior Member,~IEEE},  Engin Zeydan, \IEEEmembership{Senior Member, IEEE}, \\ Ferdi Kara,  \IEEEmembership{Senior Member, IEEE},   Esra Aycan Beyazit,  Sezai Taskin, Halim Yanikomeroglu, \IEEEmembership{Fellow, IEEE} 
\thanks{B. Karaman, I. Basturk, and S. Taskin are with Manisa Celal Bayar University, Turkiye (e-mails: \{bilal.karaman, ilhan.basturk, sezai.taskin\}@cbu.edu.tr).}%
\thanks{E. Zeydan is with Centre Tecnològic de Telecomunicacions de Catalunya (CTTC), 08860 Barcelona, Spain (e-mail: ezeydan@cttc.es).}%
\thanks{F. Kara was with the Department of Computer Engineering, Zonguldak Bulent Ecevit University, Zonguldak, 67100 Turkiye and now with Ericsson Research, Lund, 223 62, Sweden (e-mail: ferdi.kara@ericsson.com).}%
\thanks{E. A. Beyazıt is with the IDLab Research Group, University of Antwerp–IMEC, Belgium (e-mail: esra.aycanbeyazit@imec.be).}%
\thanks{H. Yanikomeroglu is with the Department of Systems and Computer Engineering, Carleton University, Ottawa, ON, K1S 5B6, Canada (e-mail: halim@sce.carleton.ca).}%
}

\maketitle

\begin{abstract}

Reliable and resilient communication is essential for disaster recovery and emergency response, yet terrestrial infrastructure often fails during large-scale natural disasters. This paper proposes a High-Altitude Platform Station (HAPS) and Reconfigurable Intelligent Surfaces (RIS)-assisted Internet of Things (IoT) communication system to restore connectivity in disaster-affected areas. Distributed IoT sensors collect critical environmental data and forward it to nearby gateways via short-range links, while the HAPS-RIS system provides backhaul to these gateways. To overcome the severe double path loss of passive RIS at high altitudes, we propose a dynamically adjustable sub-connected active RIS architecture that can reconfigure the number of elements connected to each power amplifier through switching mechanisms. Simulation results demonstrate substantial gains in downlink and uplink data rates, as well as system energy efficiency, compared with conventional passive RIS schemes. Moreover, a 1 dB increase in ground-station transmit power yields approximately 20–30 Mbps improvement in gateway data rates. These findings confirm that HAPS-RIS technology offers an effective and energy-efficient approach for resilient IoT backhaul in 6G non-terrestrial networks, \textcolor{black}{particularly in line-of-sight(LoS)-dominant HAPS--ground backhaul scenarios.}

\end{abstract}




\section{Introduction}



Natural disasters such as earthquakes, floods, wildfires, and hurricanes often cause severe damage to terrestrial communication infrastructure. Recent events \textcolor{black}{such as} the 2021 Haiti earthquake, Hurricane Ian in 2022, the 2023 Hawaii wildfires, the 2023 Kahramanmaras earthquakes in Turkiye, and large-scale floods in Malaysia in 2025 can  \textcolor{black}{serve} as examples \cite{karaman2025solutions}. Base stations collapse, optical backhaul links are severed, and power grids fail, cutting off affected regions from vital information exchange. In these critical moments, rapidly deployable and resilient communication systems are essential for situational awareness, resource coordination, and life-saving operations. The \ac{IoT} has become an essential tool in disaster management by enabling real-time environmental monitoring and intelligent decision support \cite{matracia2022post}. Distributed IoT sensors deployed across affected regions can detect and report key parameters such as gas leaks, water contamination, radiation levels, and structural instability,  \textcolor{black}{helping} authorities  \textcolor{black}{assess} risks and  \textcolor{black}{prioritize} response actions \cite{mowla2025iot}. These sensors typically communicate via short-range links, such as LoRa, ZigBee,  \textcolor{black}{or} Wi-Fi, with a nearby IoT gateway  \textcolor{black}{that} aggregates local data before forwarding it to command centers or cloud platforms \cite{karaman2025solutions}. However, when the terrestrial backhaul network is damaged or destroyed, these gateways become isolated, and the valuable data collected from the field cannot reach decision makers. Ensuring reliable backhaul connectivity for IoT gateways in  \textcolor{black}{these} situations remains a significant challenge.

Traditional solutions such as satellite links or mobile base stations face high latency, limited capacity, and complex deployment. Furthermore, the unpredictable disaster environment  \textcolor{black}{causes} severe channel impairments, power shortages, and congestion as numerous IoT nodes attempt to communicate simultaneously. \textcolor{black}{Disaster backhaul systems are fundamentally limited by restricted coverage scalability, strict energy constraints, and insufficient adaptability to rapidly changing post-disaster conditions. Addressing these limitations requires backhaul architectures that jointly optimize coverage flexibility, energy efficiency, and reconfigurability.} To address these challenges, researchers are increasingly investigating \ac{NTN} that integrate aerial communication layers with terrestrial IoT infrastructures \cite{liu2024empowering}. In this context, the combination of \ac{HAPS} and \ac{RIS} has attracted  \textcolor{black}{significant} attention \cite{matracia2024unleashing}. HAPS, positioned in the stratosphere at altitudes of 17–25  \textcolor{black}{kilometers}, can serve as aerial base stations, providing wide coverage and low-latency connectivity \cite{safwanlinkbudget}. Meanwhile, RIS, engineered meta-surfaces capable of dynamically adjusting the amplitude and phase of reflected signals, enhance wireless links by mitigating blockages and improving propagation conditions \cite{abbasi2024haps}.

\begin{figure*}[htp!]
    \centering
    \includegraphics[width=0.8\linewidth]{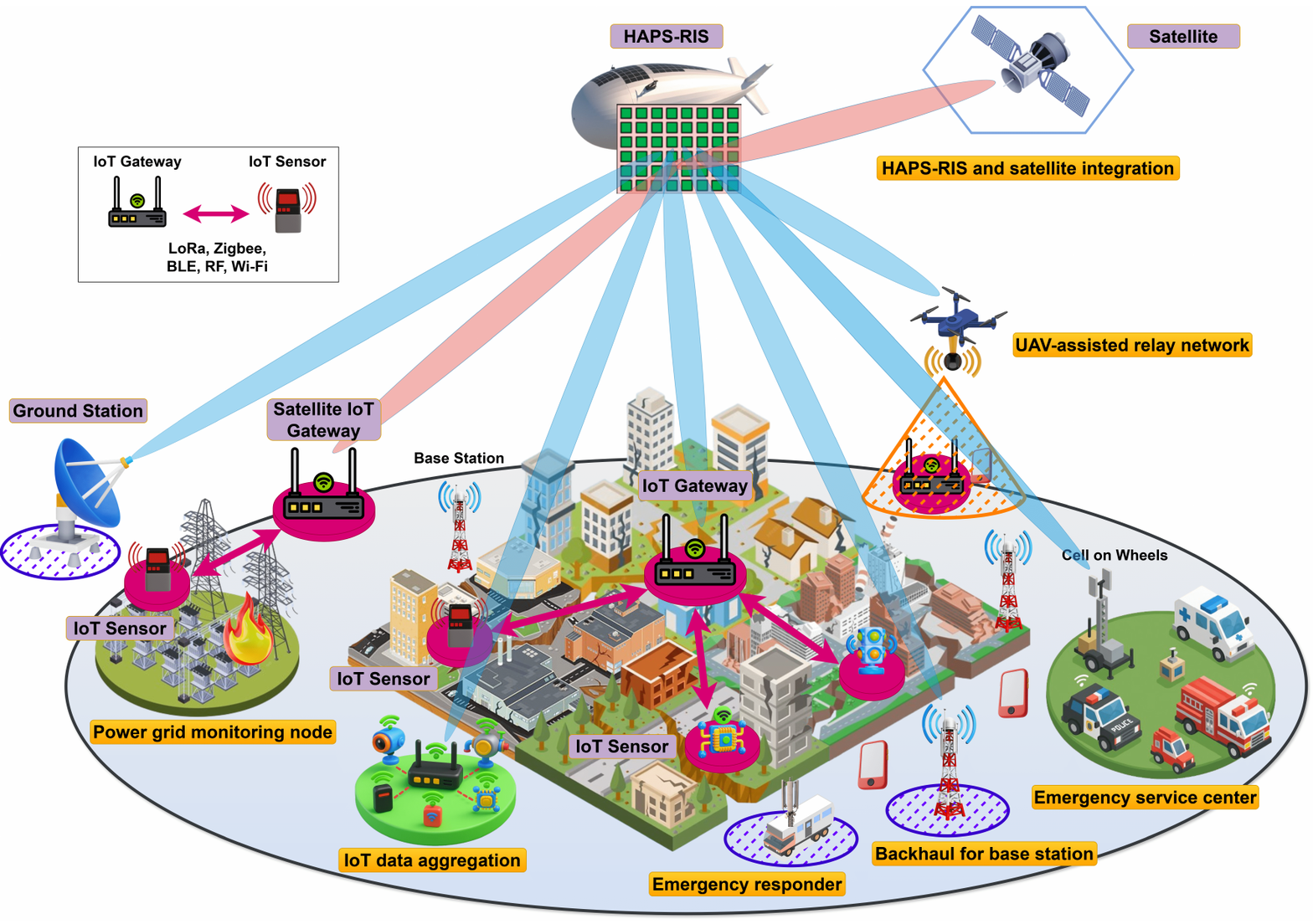}
    \caption{HAPS-RIS-assisted IoT communication system for disaster recovery and emergency response.}
    \label{fig:Fig1}
    \vspace{-.5cm}
\end{figure*}

In the proposed system shown in Fig. \ref{fig:Fig1}, IoT sensors communicate with ground-based gateways, and a HAPS-RIS-assisted network provides the essential backhaul connection between these gateways and remote control centers. The RIS elements are deployed passively or actively to improve link reliability and spectral efficiency. Passive RIS configurations operate with minimal energy consumption by reflecting incoming signals, while active RIS can amplify weak signals and mitigate severe path loss, but  \textcolor{black}{this comes} at the cost of higher power usage \cite{karaman2025trade}. To address this trade-off, we propose a dynamically adjustable sub-connected active RIS architecture that can reconfigure the number of RIS elements connected to each power amplifier  \textcolor{black}{using} switching mechanisms. This design maintains high spectral efficiency  \textcolor{black}{and} significantly  \textcolor{black}{improves} overall system energy efficiency. \textcolor{black}{Unlike conventional terrestrial RIS deployments, the wide-area coverage capability of HAPS platforms enables the integration of thousands of RIS elements, making reconfigurable sub-connected architectures practically feasible. In this context, dynamic switching among RIS element groups is not just a parameterized extension but a necessary design dimension to address the resulting energy and hardware scalability challenges. At the same time, element-level RIS phase shifts are adjusted to track time-varying channel conditions at the receivers, such as  gateway mobility, UAV relay movement, or channel fluctuations. This work leverages the HAPS-specific operating regime to reveal energy–performance trade-offs and to design a disaster-oriented RIS-assisted backhaul architecture that adapts to varying coverage and endurance requirements.} In our simulations, we evaluate the achievable downlink and uplink sum rates of IoT gateways and analyze energy efficiency performance across different RIS configurations. The results show that HAPS-RIS-assisted IoT networks can effectively restore communication in disaster-affected regions, achieving a favorable balance  \textcolor{black}{of} throughput, reliability, and energy consumption. This paper highlights the potential of integrating HAPS and RIS technologies to establish energy-efficient, adaptive, and resilient IoT communication systems for disaster recovery and emergency response.  \textcolor{black}{This is} a key step toward next-generation 6G non-terrestrial networks, \textcolor{black}{particularly in LoS-dominant HAPS--ground backhaul scenarios.} Compared to existing studies on IoT connectivity and post-disaster communication, the main contributions of this paper are as follows:
\begin{itemize}
\item We propose a HAPS–RIS-assisted IoT backhaul architecture capable of restoring connectivity across a 50 km radius disaster zone  \textcolor{black}{by} leveraging the large footprint and persistent line-of-sight (LoS) characteristics of stratospheric platforms.

\item We introduce a dynamically adjustable sub-connected active RIS design that mitigates the severe double path loss of high-altitude links. Numerical results show that reducing the group size to $L=500$ yields substantial performance gains, with gateway downlink rates improving by up to $2\times$ or $3\times$ compared to passive RIS. Furthermore, a 1~dB increase in ground-station transmit power provides an additional $20-30$~Mbps improvement in gateway data rates. We also evaluate the system's energy efficiency and show that sub-connected active RIS schemes outperform passive RIS despite higher power consumption, due to significantly higher achievable throughput.

\item We outline key deployment challenges and open research directions for practical HAPS-RIS systems in disaster recovery applications.
\end{itemize}

\textcolor{black}{\section{Enabling Technologies and Application Scenarios}}
\label{background_usecases}

This section outlines the key technologies that enable the proposed HAPS–RIS-assisted IoT communication architecture and describes representative application scenarios in both daily operations and disaster environments. The enabling technologies provide the foundation for reliable long-range connectivity, while the application scenarios illustrate how the integrated system can support diverse IoT services and restore communication capabilities when terrestrial networks are impaired.

\subsection{Enabling Technologies}

\textit{High-Altitude Platform Stations (HAPS):} \ac{HAPS} are stratospheric aerial platforms (17--25 km altitude) capable of providing large-area, low-latency coverage through persistent LoS links. Their rapid deployability and long endurance make them suitable for restoring connectivity after disasters. Ongoing International Telecommunication Union (ITU) standardization and World Radiocommunication Conference 2023 (WRC-23) decisions identifying the S-band for HAPS use support their integration into future 6G NTN systems~\cite{WP5D,wrcReport}. In disrupted regions, HAPS can act as aerial backhaul nodes, linking isolated \ac{IoT} gateways with command centers. \textcolor{black}{Compared to satellite-assisted NTN solutions, HAPS--RIS architectures offer distinct practical advantages in disaster scenarios, including significantly lower propagation latency due to their lower altitude, more favorable channel conditions, and improved energy efficiency through adaptive RIS configurations. These features make HAPS--RIS particularly suitable for latency- and capacity-critical phases of disaster recovery, where satellite systems may be constrained by long delays, limited throughput, or inflexible resource allocation.}


\textit{Reconfigurable Intelligent Surfaces (RIS):} RIS manipulate incident electromagnetic waves to enhance coverage, reduce interference, and mitigate blockage. 
{\color{black}
From a conceptual perspective, RIS architectures can be categorized as follows:
\begin{itemize}
    \item \textbf{Passive RIS:} Reflect incident signals using tunable phase shifts without amplification, resulting in minimal power consumption but limited performance under severe path loss.
    \item \textbf{Fully Connected Active RIS:} Employ per-element amplification to compensate for channel attenuation, providing strong performance gains at the cost of high power consumption and hardware complexity.
    \item \textbf{Sub-Connected Active RIS:} Enable signal amplification through shared power amplifiers serving groups of RIS elements, while preserving element-level phase control. The grouping parameter $L$ determines the number of RIS elements associated with each power amplifier, thereby enabling scalable energy--performance trade-offs.
\end{itemize}
}
Among these architectures, sub-connected active RIS offer a balanced operating point, making them particularly suitable for large-scale and power-constrained HAPS platforms \cite{karaman2025trade}.

\textit{IoT Sensing and Gateways:} Distributed IoT sensors collect critical environmental and structural information and forward it to local gateways over short-range links, such as LoRa, ZigBee, BLE, Wi-Fi \cite{matracia2022post}. These gateways require stable backhaul to deliver data to the cloud or emergency centers. HAPS–RIS integration ensures reliable data delivery even when terrestrial networks are partially or fully destroyed.

\subsection{Application Scenarios}

\textit{Daily IoT Operations:}  
(i) \textit{Agriculture:} HAPS-RIS provides wide-area connectivity for distributed agricultural sensors in regions lacking terrestrial coverage \cite{kurt2021vision}.  
(ii) \textit{Industrial/Power-Grid Monitoring:} 
Remote industrial sites, power transmission lines, and transformers rely on fixed sensor networks for safety and operational monitoring. HAPS-RIS can provide a backhaul for these sensor networks.
(iii) \textit{Smart Cities:} Smart meters, cameras, and environmental sensors gain a backup layer during outages or congestion.  
(iv) \textit{Satellite/Edge Integration:} HAPS-RIS can serve as an intermediate layer between Low Earth Orbit/Medium Earth Orbit satellites and ground IoT systems, improving reliability~\cite{abbasi2024haps}.

\textit{Disaster-Oriented IoT Operations:}  
(i) \textit{Emergency Backhaul Restoration:} When terrestrial base stations or fiber links fail, HAPS-RIS can rapidly re-establish wide-area connectivity by steering beams toward Cells-on-Wheels (CoWs), emergency vehicles, or temporary command units \cite{kurt2021vision}.  
(ii) \textit{IoT Data Aggregation:} Sensors deployed across disaster zones report vital information (gas leakage, structural stability, smoke levels) \cite{mowla2025iot}. HAPS-RIS strengthens uplink connections from multiple gateways and forwards data to cloud platforms.  
(iii) \textit{Emergency Hubs:} Temporary field hospitals or coordination centers can receive broadband connectivity even when nearby cellular infrastructure is damaged.  
(iv) \textit{Uncrewed Aerial Vehicle (UAV)-Aided Relay:} In obstructed or remote regions, UAVs equipped with relay modules can form multi-hop links between ground IoT nodes and the HAPS, ensuring full coverage without relying on surviving terrestrial infrastructure \cite{mowla2025iot}.

\section{Proposed Scenario and Methodology}

The proposed system, shown in Fig. \ref{fig:Fig1}, establishes a resilient HAPS-RIS-assisted IoT communication network for disaster recovery scenarios where terrestrial infrastructure is impaired. On the ground, numerous IoT sensors are deployed to monitor critical environmental and structural parameters, such as gas leakage, temperature variations, and infrastructure stability. These sensors communicate with nearby IoT gateways using short-range protocols (LoRa, ZigBee, BLE, or Wi-Fi). Each gateway aggregates local data and transmits it to the aerial layer, where a \ac{HAPS} operates as an aerial base station at stratospheric altitudes. The HAPS maintains backhaul links between the IoT gateways and a remote control center, and can also cooperate with satellites or UAV relays to ensure full coverage of the disaster area.

\begin{figure}[htp!]
    \centering
    \includegraphics[width=0.84\linewidth]{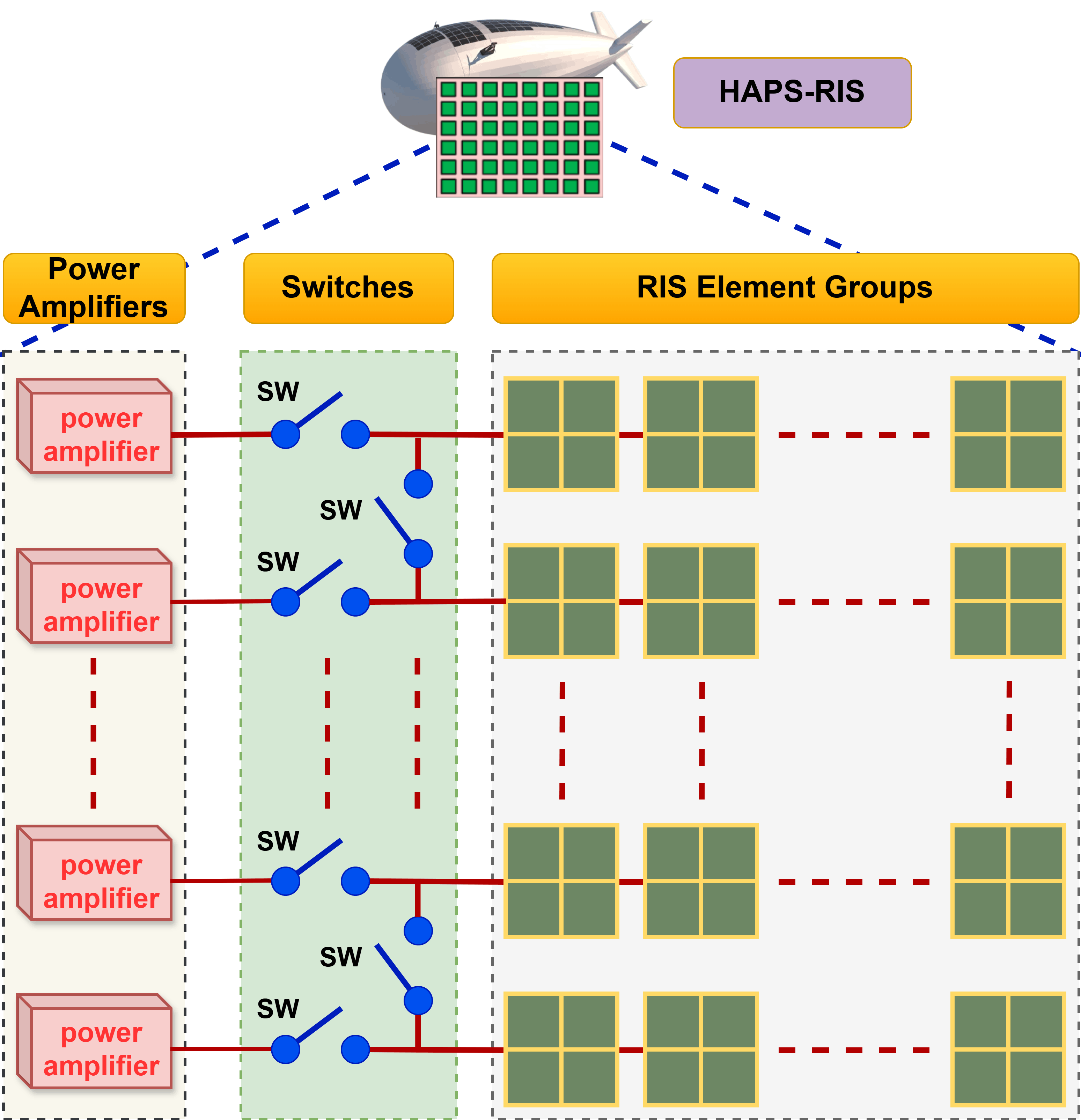}
    \caption{Dynamically adjustable sub-connected active RIS architecture.}
    \label{fig:Fig2}
    \vspace{-16pt}
\end{figure}

Given the long distances between the ground station and the HAPS, as well as between the HAPS and terrestrial devices, passive RIS systems experience significant double-fading losses. To address this limitation, active RIS architectures incorporate power amplifiers that apply controllable phase shifts after signal amplification. To reduce overall power consumption and hardware complexity in large-scale active RISs, a sub-connected configuration can be used, where each amplifier is shared among a group of $L$ elements \cite{karaman2025trade}. While the amplification factor is common within each group, each element still maintains independent phase-shift control. Considering the massive number of RIS elements that can be mounted on a HAPS surface, we propose a dynamically adjustable sub-connected active RIS architecture, in which the number of elements connected to each power amplifier ($L$) can be adaptively reconfigured through switching mechanisms. Moreover, in the proposed design, all power amplifiers can be switched off, allowing the system to operate entirely as a passive RIS and thereby achieve energy savings. Depending on the specific application scenario, this flexible grouping strategy can be configured to optimize sum-rate performance, enhance energy efficiency, or minimize power consumption. The proposed architecture is illustrated in Fig.~\ref{fig:Fig2}.

\section{Case Study}

In this section, the communication performance of the proposed HAPS-RIS-assisted IoT communication system is evaluated for a disaster-affected region. Simulation results illustrate the achievable downlink and uplink sum rates of IoT gateways, along with the corresponding system energy efficiency. \textcolor{black}{The simulations rely on standardized 3GPP air-to-ground channel models that incorporate elevation-angle-dependent LoS/NLoS probabilities, large-scale path loss, shadowing, and atmospheric attenuation \cite{safwanlinkbudget, 3GPP_UE}.} \textcolor{black}{Perfect channel state information (CSI) and ideal RIS phase-shift control are assumed in order to focus on fundamental system-level trade-offs.} Atmospheric impairments such as rain attenuation and scintillation are not explicitly modeled, as S-band links are generally more resilient to such effects than higher-frequency bands. \textcolor{black}{According to ITU-R P.618, atmospheric attenuation at S-band is typically small (on the order of $1$ dB or less) and is therefore unlikely to significantly affect the comparative trends analyzed in this study \cite{ITU_P618_14}.} The overall simulation parameters are provided in Table~\ref{tab:sims}. 

\textcolor{black}{In order to assess the backhaul capabilities of the HAPS--RIS-assisted communication system shown in Fig.~\ref{fig:Fig1}, this study concentrates on RIS-enabled HAPS backhaul architectures, leveraging the shorter ground-to-HAPS link distance relative to satellite-based systems.}
\textcolor{black}{It is worth noting that the analysis focuses on system-level performance and therefore does not assume a specific disaster traffic model or service-level QoS profile. Instead, the reported data rate and energy-efficiency distributions are intended to provide generic performance results that can support diverse emergency services, ranging from low-rate sensing and voice communication to higher-rate applications such as situational video transmission.} In the simulations, the HAPS is located at the center of a circular region with a $50$~km radius and an altitude of $20$~km. In this scenario, the ground station is positioned at coordinates ($5$~km, $5$~km) within a disaster-resilient infrastructure equipped with a high-gain antenna. This ground station serves as a communication and control hub, enabling the beam steering of RIS elements toward target IoT gateways, UAVs, CoWs, or emergency response units.

\begin{table}[htp!]
\centering
\caption{Main simulation parameters.}
\renewcommand{\arraystretch}{1.2}
\setlength{\tabcolsep}{7pt}
\begin{tabular}{|l|l|}
\hline
\rowcolor[gray]{0.9}
\textbf{Parameter} & \textbf{Value} \\ \hline
Frequency & $2.4~\mathrm{GHz}$ \\ \hline
HAPS altitude & $20~\mathrm{km}$ \\ \hline
Radius of disaster area & $50~\mathrm{km}$ \\ \hline
Ground station antenna gain & $43.2~\mathrm{dBi}$ \\ \hline
IoT gateway antenna gain & $0~\mathrm{dBi}$ \\ \hline
Signal bandwidth & $100~\mathrm{MHz}$ \\ \hline
\textcolor{black}{Number of RIS elements ($N_{\mathrm{total}}$)} & \textcolor{black}{$30000$} \\ \hline
\textcolor{black}{RIS element phase-shift power ($P_{\mathrm{sw}}$)} & \textcolor{black}{$7.8~\mathrm{mW}$} \\ \hline
\textcolor{black}{RIS element biasing power ($P_{\mathrm{dc}}$)} & \textcolor{black}{$-5~\mathrm{dBm}$} \\ \hline
Power amplifier output power ($P_{\mathrm{A}}$) & $2~\mathrm{W}$ \cite{karaman2025trade} \\ \hline
Path loss and channel models & Based on \cite{3GPP_UE}, \cite{safwanlinkbudget} \\ \hline
Noise spectral density & $-174~\mathrm{dBm/Hz}$ \\ \hline
\end{tabular}
\label{tab:sims}
\end{table}

\begin{figure}[htp!]
    \centering
    \includegraphics[width=0.85\linewidth]
    {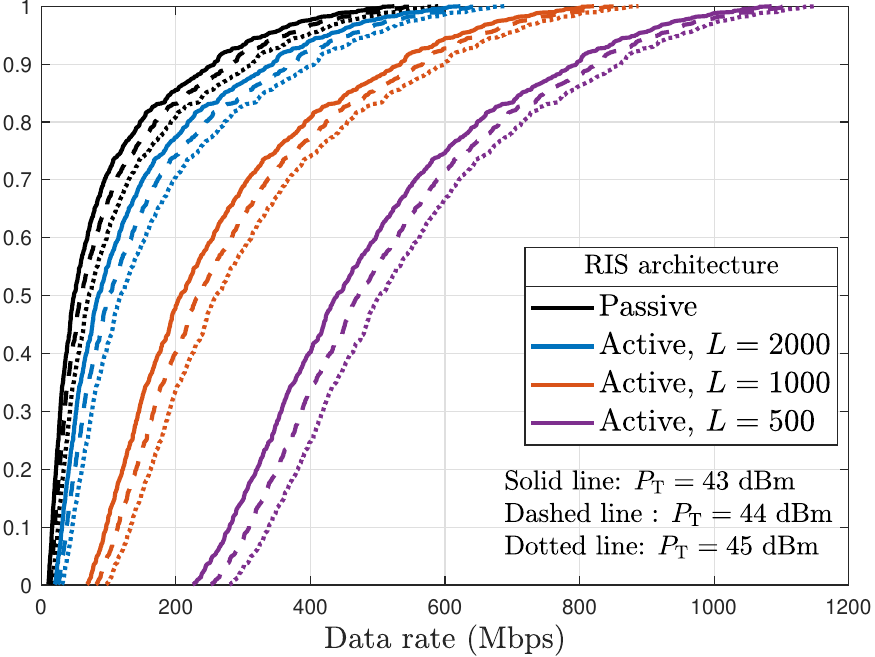}
    \caption{Cumulative distribution function (CDF) of the data rates achieved by the IoT gateways in downlink communication.}
    \label{fig:Fig3}
    \vspace{-8pt}
\end{figure}

In Fig.~\ref{fig:Fig3}, the downlink communication between the ground station and the IoT gateways via HAPS-RIS operates in the S-band (2.4~GHz). In addition to the passive RIS configuration, the sub-connected active RIS architecture is evaluated under three different schemes, where each power amplifier is shared among $L$ elements ($L=2000$, $L=1000$, and $L=500$). For the downlink communication simulations, an urban scenario is considered in which $1000$ IoT gateways are randomly distributed within a disaster area. The HAPS-RIS system comprises $30000$ RIS elements. 
\textcolor{black}{As illustrated in Fig.~3, in the passive RIS case, approximately 70\% of the gateways experience downlink data rates below 70~Mbps, whereas in the $L=1000$ configuration, even the worst-performing gateway achieves at least 70~Mbps.}The results show that sub-connected active RIS schemes outperform the passive RIS configuration in achievable sum rate. As expected, reducing the group size $L$ increases the number of power amplifiers, which consequently improves the sum-rate performance of the IoT gateways. \textcolor{black}{Moreover, as shown in Fig.~\ref{fig:Fig3}, the transmit power of the ground station has a dominant influence on overall system performance. Under the considered simulation parameters, a 1~dB transmit-power increase yields an observed rate gain of approximately 20--30~Mbps at the IoT gateways.} If the HAPS-RIS system is intended to provide backhaul connectivity for terrestrial base stations, emergency responders, UAVs, or CoWs, high-gain receiver antennas (e.g., 15 dBi) can be mounted on these platforms to enhance link quality and achievable capacity.

\begin{figure}[htp!]
    \centering
    \includegraphics[width=0.85\linewidth]
    {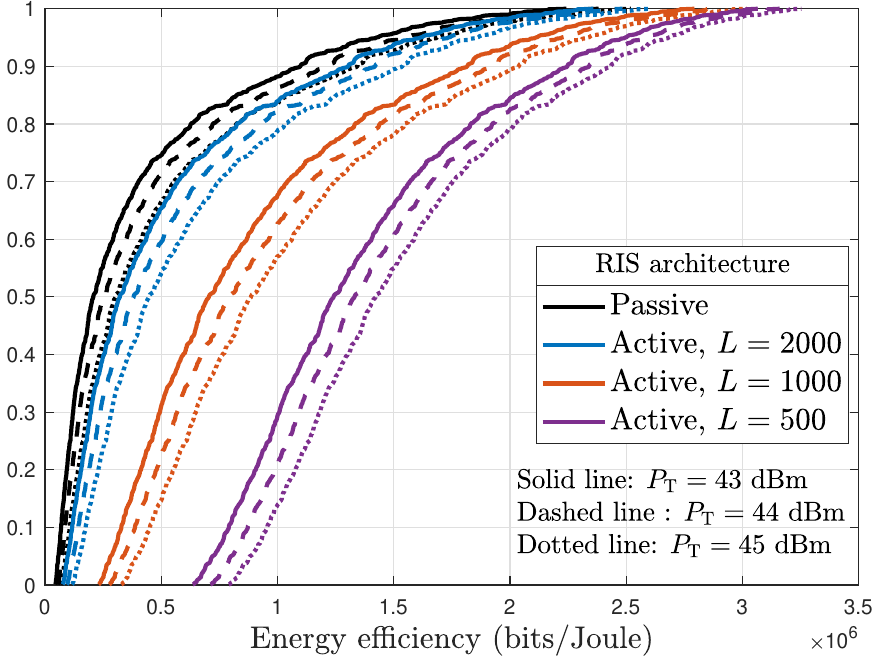}
    \caption{CDF of the energy efficiency at the IoT gateways for downlink communication.}
    \label{fig:Fig4}
    \vspace{-8pt}
\end{figure}

Fig.~\ref{fig:Fig4}, in contrast, presents the energy efficiency comparison between sub-connected active and passive RIS architectures. \textcolor{black}{The total power consumption at the RIS is modeled by accounting for the dominant communication-related components. Specifically, the total RIS power consumption is expressed as $P_{\mathrm{RIS}} = N_{\mathrm{total}} P_{\mathrm{sw}} + N_{\mathrm{total}} P_{\mathrm{dc}} + (N_{\mathrm{total}}/{L}) P_{\mathrm{A}}$,
where $P_{\mathrm{sw}}$ denotes the power required for phase shifting of each RIS element, $P_{\mathrm{dc}}$ represents the biasing power consumed by each RIS element, and $P_{\mathrm{A}}$ is the power consumed by each active power amplifier. Here, $N_{\mathrm{total}}$ is the total number of RIS elements, and the term $N_{\mathrm{total}}/L$ corresponds to the number of active power amplifiers in the sub-connected architecture \cite{karaman2025trade}.} Although the sub-connected active RIS consumes more power, it achieves higher energy efficiency due to the significant improvement in sum-rate performance. \textcolor{black}{From Fig.~4, it is observed that nearly 80\% of the gateways in the passive RIS case operate below an energy efficiency of 0.62~Mbit/Joule, while in the $L=1000$ configuration all gateways achieve energy efficiency values above this threshold.} However, a smaller $L$ value requires a larger number of power amplifiers (for example, when $L=2000$, a total of 15 amplifiers are needed, whereas for $L=500$, 60 amplifiers are required). The increase in the number of amplifiers inevitably raises the total energy consumption at the RIS. As a result, adopting a fully connected active RIS configuration would lead to poor energy efficiency. Given the limited power budget of the HAPS, employing sub-connected active RIS schemes with fewer power amplifiers is expected to extend operational duration and ensure sustainable performance.

\begin{figure}[htp!]
    \centering
    \includegraphics[width=0.85\linewidth]
    {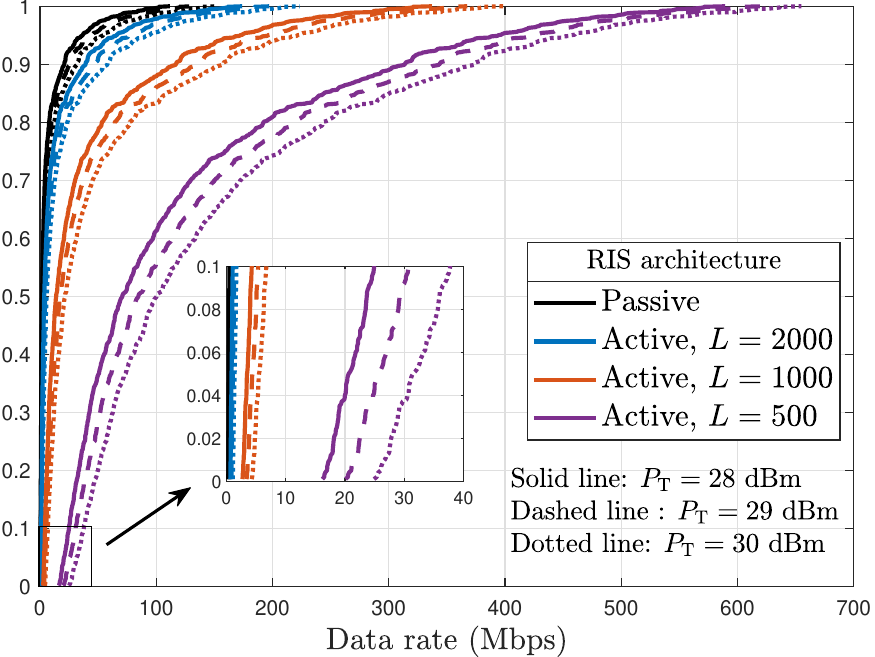}
    \caption{CDF of the uplink data rates of the IoT gateways.}
    \label{fig:Fig5}
    \vspace{-5pt}
\end{figure}

In Fig.~\ref{fig:Fig5}, the uplink communication between the IoT gateways and the ground station via the HAPS-RIS is evaluated. The transmission of data collected from IoT sensors in the disaster area by IoT gateways is crucial for emergency response and coordination. However, as IoT gateways are typically battery-powered and low-power devices, their transmit power is assumed to be in the range of $28$-$30$ dBm. Under these conditions, the sub-connected active RIS configuration with $L=500$ achieves the highest performance, providing data rates between $15$ Mbps and $650$ Mbps. \textcolor{black}{As shown in Fig.~ 5, about 85\% of the gateways in the passive RIS scenario achieve uplink data rates below 15 ~Mbps, whereas in the $L=500$ case, the minimum uplink data rate across all gateways exceeds 15 ~Mbps.} \textcolor{black}{The CDF gains result from active RIS signal amplification and increased array gain at smaller $L$ due to a higher number of active power amplifiers, and are most pronounced in power-limited uplink and long-distance HAPS–ground links.}

From these results, the following conclusions can be drawn: \textcolor{black}{For downlink communication, increasing the transmit power of the ground station and employing a passive RIS can reduce the overall system energy consumption, resulting in power savings at the HAPS platform. In contrast, for uplink communication, where IoT gateways are typically power-limited, switching to a sub-connected active RIS architecture can significantly improve the achievable data rates and link reliability. The numerical results reveal a clear performance–energy trade-off controlled by the RIS grouping parameter $L$. Smaller values of $L$, corresponding to a larger number of active power amplifiers, maximize achievable throughput and link robustness, whereas larger $L$ values reduce total power consumption and improve HAPS endurance. These insights indicate that passive RIS is preferable in downlink-dominated, energy-constrained operation, while active RIS with appropriately selected $L$ should be employed when uplink performance or coverage reliability is critical.} 
\textcolor{black}{The considered deployment of $30000$ RIS elements represents an early-stage system-level estimate intended to illustrate the scalability of HAPS-assisted RIS architectures. Assuming a unit-cell size of $(0.2\lambda)^2$ as in~\cite{kurt2021vision}, the total surface area required by $30000$ elements is approximately $27~\mathrm{m}^2$, which constitutes a limited fraction of a typical HAPS surface. Moreover, considering the weight of each RIS element to be $10~\mathrm{g}$ as commonly assumed in analytical and simulation-based studies such as~\cite{tyrovolas2022energy, matracia2024unleashing}, the total payload mass associated with the RIS amounts to approximately $300$ kg, highlighting the need to account for practical integration and payload constraints in future hardware-oriented studies.}
\textcolor{black}{Using the adopted power model, for $N_{\mathrm{total}}=30000$ and $L=500$ (i.e., $60$ amplifiers) with the parameters in Table~1, the total RIS power consumption is approximately $P_{\mathrm{RIS}}\approx 364~\mathrm{W}$, which is presented as an early-stage feasibility estimate that neglects conversion losses and energy storage requirements. Based on current solar panel technology, this power level can be supplied by solar panels occupying on the order of $1~\mathrm{m}^2$ of surface area under favorable conditions~\cite{rosabal2022minimization}.}

\textcolor{black}{To further contextualize the benefits of RIS assistance, the performance gain over a direct HAPS-to-ground link depends strongly on the operating regime. In scenarios with moderate elevation angles and favorable line-of-sight conditions, a direct HAPS link may provide sufficient link margin without the additional complexity of RIS. However, RIS assistance becomes advantageous in low-elevation or blockage-prone environments, where passive beam shaping enhances the received signal strength without increasing transmit power. Compared to satellite backhaul, HAPS-RIS configurations offer lower latency and reduced slant-path loss, while satellite systems remain advantageous for wide-area or global coverage.}

\section{Challenges and Future Directions}

In this section, we discuss the main practical challenges associated with deploying HAPS-RIS-assisted IoT networks in disaster scenarios, as well as promising directions for future research. While our numerical results demonstrate the potential benefits of the proposed architecture, several implementation, control, and environmental issues must be addressed before large-scale deployment becomes feasible.

\subsection{Challenges}

We first outline key technical and operational challenges that may hinder the realization of HAPS-RIS-assisted IoT systems. These challenges provide a roadmap for future research and guide the design of more robust and scalable solutions.

\textit{Real-time phase configuration:} Real-time configuration of RIS elements is challenging, particularly when thousands of elements are mounted on a HAPS platform. While communication with fixed ground devices, such as terrestrial base stations or distributed IoT gateway nodes, typically exhibits slow channel variations, mobility-intensive scenarios introduce rapid fluctuations in channel conditions. UAVs, emergency vehicles, or moving responders can cause fast channel dynamics that demand frequent and precise phase updates. Moreover, the long ground device-HAPS distance introduces feedback delays and increases the difficulty of obtaining accurate CSI. 
\textcolor{black}{Although sub-connected active RIS architectures do not reduce the dimensionality of phase control, they can indirectly alleviate real-time control challenges by improving channel estimation quality. By amplifying the incident signal, the RIS enhances pilot SNR and feedback reliability, which is especially critical for long-distance and power-limited HAPS–ground links. This improved CSI quality reduces sensitivity to feedback errors and delays, contributing to more stable real-time RIS operations.
}

\textit{Vibration:} Although HAPS platforms experience far fewer vibrations than UAV-mounted RIS systems, residual mechanical oscillations and platform drift can still disturb the RIS orientation and reflected beam direction. Even small angular deviations may degrade the alignment toward distant IoT gateways because of the high sensitivity of RIS-assisted LoS paths. These fluctuations complicate channel estimation and limit the achievable beamforming accuracy, especially when operating at high frequencies. \textcolor{black}{This challenge can be mitigated by using mechanically stabilized RIS mounting combined with sensor-assisted orientation control to maintain stable RIS alignment.}

\textit{Power consumption:} Active RIS architectures can improve link quality by amplifying weak HAPS-ground signals, but they require additional power for their amplifiers and control circuitry. In disaster scenarios, available energy on the HAPS platform is constrained by solar harvesting and battery reserves, making continuous active operation difficult. Balancing amplification gains with energy limitations is therefore essential. \textcolor{black}{This challenge can be addressed by adaptively controlling the grouping parameter~$L$ based on real-time energy availability and traffic demand, enabling energy-aware operation of sub-connected active RIS.}

\textit{Deployment:} Deploying HAPS-RIS systems for disaster recovery and emergency response presents several practical considerations. While HAPS platforms can be rapidly dispatched to disaster regions without requiring fibre infrastructure or additional terrestrial equipment, the integration of large RIS units slightly increases payload and maintenance needs. Nevertheless, routine retrieval of HAPS platforms can facilitate periodic inspection, calibration, and upgrades of the RIS units. Once airborne, the system can quickly establish backhaul links and restore connectivity across wide areas. Remaining challenges primarily relate to optimizing RIS alignment and control under dynamic atmospheric conditions, which continue to be active research problems rather than fundamental deployment barriers. \textcolor{black}{These challenges can be alleviated through modular RIS integration, periodic ground-based calibration during HAPS retrieval cycles, and adaptive alignment control to maintain robust operation under time-varying atmospheric conditions.}

\subsection{Future Directions}

We now highlight several promising future research directions for HAPS-RIS-assisted IoT networks. These directions aim to bridge the gap between theoretical performance gains and practical deployment, and can guide the development of more intelligent, energy-aware, and resilient non-terrestrial communication architectures.

\textit{Artificial Intelligence (AI)/Machine Learning (ML)-Driven RIS Configuration and Resource Optimization:} The dynamic nature of disaster scenarios makes real-time optimization of RIS parameters challenging. Future research can explore advanced AI and ML techniques for beamforming design, RIS grouping, phase-shift control, channel estimation, and resource allocation. Learning-based approaches, especially reinforcement learning and graph neural networks, can enable fast adaptation to rapidly changing conditions while reducing signaling overhead. Moreover, lightweight ML models tailored to the computational constraints of NTN platforms will be crucial for practical deployment.

\textit{Practical RIS Control with Group-Based or Hierarchical Architectures:} Configuring thousands of RIS elements individually is impractical in real-world systems because of hardware limitations, signaling overhead, and control latency. A promising approach is to explore group-based, hierarchical, or hybrid control mechanisms, where subsets of RIS elements share a common phase or amplification configuration. These architectures can greatly reduce control complexity while maintaining high reflection efficiency. Developing optimal grouping strategies that adapt to channel dynamics and energy constraints remains an open research problem.

\textit{Joint Communication-Energy Optimization Under HAPS Constraints:} HAPS platforms operate under strict payload, power, and endurance limitations. Future efforts should consider joint communication-energy optimization, including amplifier scheduling, dynamic switching between passive and active RIS modes, sleep-mode operation for RIS groups, and energy-aware gateway prioritization. Integrating real-time power forecasting based on expected solar energy availability into communication scheduling can improve network longevity during prolonged disaster scenarios.

\textit{Multi-HAPS Cooperation and Satellite-IoT Integration:}
Future HAPS-RIS deployments can benefit from multi-layer NTN architectures in which multiple HAPS platforms cooperate to enhance coverage, capacity, and reliability. Coordinated beamforming, inter-HAPS handover management, and distributed control of multiple RIS units will be essential for large-scale disaster areas. Additionally, integrating HAPS-RIS systems with existing satellite-IoT networks, already widely deployed for remote monitoring, can enable hybrid routing strategies and seamless data offloading. Such multi-tier architectures can ensure uninterrupted IoT connectivity even when individual platforms experience blockage, overload, or energy shortages.

\textit{Scalable Channel Estimation and Robust Operation Under Environmental Dynamics:} Accurate channel estimation for long-range HAPS-ground devices links remains a significant challenge due to feedback delays, atmospheric variations, and limited pilot resources. Future research should develop scalable estimation schemes that minimize signaling overhead while maintaining robustness in rapidly changing environments. Additionally, environmental factors such as wind-induced platform motion, thermal fluctuations, and mechanical vibration may degrade the phase alignment of RIS elements. Designing adaptive compensation mechanisms and robust beamforming strategies to preserve link quality under these uncertainties will be critical for practical deployment.

\ifCLASSOPTIONcaptionsoff
  \newpage
\fi

\vspace{-8pt}
\section{Conclusions}

This study shows that HAPS--RIS technology has strong potential for next-generation NTN supporting disaster recovery and emergency response. Integrating HAPS with dynamically configurable RIS enhances backhaul connectivity and enables robust IoT data delivery under impaired terrestrial conditions. The proposed sub-connected active RIS design allows flexible operation modes that balance spectral efficiency and power consumption under varying link conditions.
{\color{black}
Rather than relying solely on raw performance gains, the results offer clear design insights for practical deployment. Passive RIS configurations are well suited for downlink-dominated scenarios with sufficient ground station transmit power, providing energy-efficient operation at the HAPS platform. In contrast, sub-connected active RIS is critical in uplink-limited and power-constrained settings, where signal amplification significantly improves link robustness. Moreover, the RIS grouping parameter~$L$ enables a controllable trade-off between throughput and energy consumption, allowing the system to adapt to different operational phases and endurance requirements.
}
Overall, the proposed architecture offers a resilient and energy-efficient solution for rapidly restoring communication in large-scale disaster-affected areas.

\balance


\bibliographystyle{ieeetr}
\bibliography{biblio_clean_r2}  








\vspace{.1cm}

\end{document}